\documentclass[12pt]{article}
\usepackage{authblk}
\usepackage{physics}
\usepackage{amsmath}
\begin{document}
\title{\textbf{On the collapse of states and the uncertainty principle}}
\author{Sankarshan Sahu}
\affil{Indian Institute of Engineering Science and Technology, Shibpur}
\maketitle
\begin{abstract}
There had been previous successful explanations of Quantum Mechanics using the popular Copenhagen interpretation. In this paper, we build an equivalent mathematical structure of Copenhagen Interpretation from the Electromagnetic Field Theory and show how uncertainty principle is associated with the 'collapse of states'. We also try to provide a more intuitive method towards our understanding of Quantum Mechanics.
\end{abstract}
\section*{Introduction}
There had been previous works showing how electromagnetic fields can be explained using coherent states[1-4].There also had been notable works relating gravity with uncertainty principle[5]. Here we try to build a new explanation for Copenhagen Interpretation using Electromagnetic Field Theory in Coherent states and the uncertainty principle. We try to implement a new mathematical model which intuitively explains 'the collapse of states' which till now was postulated as a quantum mechanical phenomenon. A special case of perturbation in quantum harmonic oscillator (which also governs the Hamiltonian for electromagnetic fields in a box)
is studied. We assume the perturbed potential to
be a Harmonic Oscillator that has been shifted by a fixed amount in the position space as well as momentum space.The Perturbation is thought to arise from the only available interaction i.e. observation. \\
\\
\\
We start with the Hamiltonian as in [6], governing the Electromagnetic fields in a box, of the form
\begin{equation}
\hat{H}=\frac{1}{2}(\hat{p}^2+\omega^2 \hat{q}^2)    
\end{equation}
which is that of a Harmonic Oscillator. As shown in [7], The Electric fields show the classical behaviour i.e. follow the Maxwell's equation (standing wave solution), when the expectation value of the time dependent $\hat{p}$ is calculated in the coherent states $\ket{\alpha}$, where[7]:-
\begin{equation}
\ket{\alpha}=e^{-\frac{1}{2}|\alpha|^2}\sum_{n=0}^{\infty}\frac{\alpha^{n}}{\sqrt{n!}}\ket{n}    
\end{equation}
where $\ket{n}$ are the energy eigenvalues of the Hamiltonian in (1).
These coherent states can also be obtained as :-
\begin{equation}
\ket{\alpha}=D(\alpha)\ket{0}   
\end{equation}
where,
\begin{equation}
D(\alpha)= exp({\alpha\hat{a}^{\dagger}-\alpha^{*}\hat{a}})
\end{equation}
where $\hat{a}$ and $\hat{a}^{\dagger}$ are the annihilation and creation operator respectively.
That is, we have:-
\begin{equation}
\bra{\alpha}\hat{E_x}(z,t)\ket{\alpha}=2|\alpha|\varepsilon_0\textbf{Re}({\alpha}e^{-i{\omega}t})\sin\textit{kz}=2\varepsilon_0|\alpha|\cos({\omega}t-\theta)\sin{\textit{kz}}
\end{equation}(The classical standing wave solution to Maxwell's equation.)
However, when the expectation value of the Electric field operator is calculated in the number states(photon states) (states that diagonalize the Hamiltonian in (1)), we have:-
\begin{equation}
\bra{n}\hat{E_x}(z,t)\ket{n}= \varepsilon_0(\bra{n}\hat{a}\ket{n}e^{-i{\omega}t}+\bra{n}\hat{a}^{\dagger}e^{i{\omega}t}\ket{n})\sin{\textit{kz}}=0
\end{equation}
This phenomenon accurately describe the double slit experiment in accordance with the Copenhagen interpretation.\\ Since $\ket{\alpha}$ can be written as in (2), one can view coherent states as a superposition of superposition of number states. Whenever a measurement is made, the coherent state$\ket{\alpha}$ collapses into one of the number states $\ket{n}$(in accordance with the Copenhagen interpretation), since the expectation value of Electric field is 0 in number states wave nature of light is not observed. However when no measurement is made, no collapse of states take place essentially and the Electric field gets some contribution from all the number states $\ket{n}$(i.e. the expectation value of the Electric field is calculated in $\ket{\alpha}$ states) and we get the classical electric field as in (5).
Going further, we will develop a new intuition in our understanding of collapse of states, and in a way replace the collapse of states with a new Mathematical model.\\ We start with the well known Schrodinger's equation:-
\begin{equation}
\hat{H}\ket{\Psi}=E\ket{\Psi}    
\end{equation}
In the Copenhagen interpretation, whenever a measurement is made we work under the assumption that a collapse of states take place instantaneously . However in this model, our focus is not to work on eigenstates i.e. we keep the eigenstates intact and work with the only other element,  the Schrodinger's equation can offer i.e. the Hamiltonian, i.e. we assume the instantaneous change in the Hamiltonian.
We basically say that observation results in the change of the Hamiltonian rather than a change in the state.
Let us give the Hamiltonian an unitary evolution.
\begin{equation}
\hat{H}_{\alpha}=D(\alpha)\hat{H}D(\alpha)^{\dagger}
\end{equation}
However the change in the Hamiltonian has to be instantaneous, since we implement the instantaneous collapse of states. We treat the new Hamiltonian as  a perturbation to our old system i.e.$\hat{H}$, such that the sudden approximation[8] can be applied.
The energy eigenstates of the Hamiltonian $\hat{H}_{\alpha}$ can thus be expressed as:-
\begin{equation}
\ket{\alpha}_{n}=D(\alpha)\ket{n}    
\end{equation}
with the coherent state $\ket{\alpha}=\ket{\alpha}_{0}$ being its ground state.
Let the Hamiltonian describing the electromagnetic field be described as:-
\
\begin{equation}
\hat{H}=
\hat{H}_{\alpha} \;at\;(t=t_1)\;and\;
\hat{H}\;at\; (t=t_{2})
\end{equation}

Thus the Energy eigenstates evolve as:-\\
$\ket{\psi}(t)=\ket{\psi}(t_{1})$ at $t_{1}<t<t_{2}$, and\\
$\ket{\psi}(t>t_{2})=e^{-iH(t-t_{2})/{\hbar}}\ket{i_{1}}$[8],
Thus the expectation value of the operator $<p(t)>$ can be shown to be:-
\begin{align}
<\hat{p}(t)>=
\bra{\psi(t_{1})}\hat{p}\ket{\psi(t_{1})} \;at\; t_{1}\leq t \leq t_{2},\\
<\hat{p(t)}>=\bra{\psi(t_{1})}e^{i H(t-t_{2})}\hat{p}e^{-i H(t-t_{2})}\ket{\psi(t_{1})} \;at\: t\geq t_{2}
\end{align}
Between $t_{1}$ and $t_{2}$ the eigenstates of  the Hamiltonian $<\hat{p}(t)>$
doesn't evolve and the value of $<\hat{p}(t_{1}<t<t_{2})>$ comes out to be zero.However since the change is assumed to be instantaneous $t_{1}$ and $t_{2}$ are almost same.
However for $t\geq t_{2}$ :-
\begin{align}
<\hat{p}(t)> & = \bra{\psi(t_{1})}e^{i H(t-t_{2})}\hat{p}e^{-i H(t-t_{2})}\ket{\psi(t_{1})} \\
 & = \bra{\psi(t_{1})}\hat{a}e^{i\omega t}+\hat{a}^{\dagger}e^{-i\omega t}\ket{\psi(t_{1})}\\
 & =\bra{\alpha}\hat{a}e^{i\omega (t-t_{2})}+\hat{a}^{\dagger}e^{-i\omega (t-t_{2})}\ket{\alpha}\\
 & = 2|\alpha|\cos{(\omega (t-t_{2}) - \theta)}
 \end{align}
 Here $\ket{\psi(t_{1})}$ is taken to be the ground state of the Hamiltonian $H_{\alpha}$ i.e. $\ket{\alpha}=D(\alpha)\ket{0}$ where $\ket{0}$ is the eigenstate of the Hamiltonian $H$.
Thus the corresponding Electric Field can be written as \begin{equation}
    <E_{x}(z,t)>=2|\alpha|\epsilon_{0}\cos{(\omega t - \theta)}\sin{kz}
\end{equation}
Similarly we can find the magnetic field to be :-
\begin{equation}
<B_{y}(z,t)>= B_{0}\sin{(\omega t - \theta)}\cos{kz}  
\end{equation}
(the factor of $\omega t_{2}$ has been absorbed in the phase difference $\theta$)
The above result and calculation holds true for any state eigenstate of the Hamiltonian $\hat{H}_{\alpha}$ and for a generalized $\ket{\alpha}_{n}$ and not only for its ground state. Thus here for obtaining the classical value of Electric field we remain in the eigenstates of the Hamiltonian $\hat{H}_{\alpha}$, and assume a sudden transition in the Hamiltonian. This describes the wave nature of light. 
Now similarly for obtaining the particle nature of light, we assume no change(transition) in Hamiltonian takes place(i.e. we again remain in the energy eigenstates of $\hat{H}_{\alpha}$ and the time evolution takes place in accordance to $\hat{H}_{\alpha}$), and thus we obtain:-
\begin{equation}
      \bra{\alpha}_{n}e^{iH_{\alpha}t}\hat{p}_{\alpha}e^{-iH_{\alpha}t}\ket{\alpha}_{n}=0 
\end{equation}
Thus we see in the new model proposed by us, act of no-measurement(non-observing) is mathematically represented by the instantaneous change in Hamiltonian where as act of measurement(observation) is mathematically represented by no change in the Hamiltonian. This is exactly opposite of what happens with the states, where the act of observation is responsible for change or collapse of states where as act of 'non-measuring' or non observing guarantees that the system stays in a superposition of states.
\\However, this new model precisely matches the results of the older one(Copenhagen Interpretation) and we will further see how at times it is more useful while explaining some fundamental relations.
Previously the number states $\ket{n}$ diagonilized the Hamiltonian $\hat{H}$ but was unable to give the picture of classical waves, for whose description we had to use coherent states $\ket{\alpha}$. In our new model, we only use one state $\hat{n}_{\alpha}$ but two Hamiltonians $\hat{H}_{\alpha}$ and $\hat{H}$ where the Transition of Hamiltonian from $\hat{H}_{\alpha} \rightarrow \hat{H}$ gives the classical Electromagnetic wave where as using just $\hat{H}_[\alpha]$ gives us the particle nature of light  .
From (8), we have:-
\begin{equation}
\hat{H}_{\alpha}=\frac{1}{2}((\hat{p}-\Delta{p} |{\alpha}|\sin{\theta})^2+{\omega}^2(\hat{q}-\Delta{q} |{\alpha}|\cos{\theta})^2)     
\end{equation}
where $\Delta p= \bra{\alpha}_{n}\hat{p}^2\ket{\alpha}_{n}-\bra{\alpha}_{n}\hat{p}\ket{\alpha}_{n}^2$, and
 $\Delta q= \bra{\alpha}_{n}\hat{q}^2\ket{\alpha}_{n}-\bra{\alpha}_{n}\hat{q}\ket{\alpha}_{n}^2 $
 Thus we can come up with the explanation, for the transition of Hamiltonian. The answer is pretty straight forward, and is what one would expect from basic intuition i.e. the observed Hamiltonian is different from the initial Hamiltonian governing the states, due to the uncertainties associated with the measurement. The Hamiltonian $H_{\alpha}$ is thus the observed Hamiltonian, since it has the terms associated with the uncertainties.
 (The values of $\Delta p(t)$ and $\Delta q(t)$ are both time-independent and thus at any point in time the uncertainties associated with canonical position and momentum remain the same, which is a signature of the coherent state )
\\ However we also have a phase factor $\theta$ and a scaling factor associated with $\alpha$. Our next task is to find a meaningful answer if we only take the raw value $\Delta p$ and $\Delta q$ by setting $\theta=\pi/4$ and $|\alpha|=\sqrt{2}$, thus we get a Hamiltonian free of any scaling factor or phase factor.
\begin{equation}
\hat{H}_{\pi/4}=\frac{1}{2}((\hat{p}-\Delta p)^2 +{\omega}^2 (\hat{q}-\Delta q)^2)
\end{equation}
Following this Hamiltonian we have:-
\begin{equation}
    <{E_x}(z,t)>= E_0\cos(\omega t - \omega t_2-\pi/4)\sin{kz}
\end{equation}
and,
\begin{equation}
    <B(t)>=B_0\sin(\omega t - \omega t_2 -\pi/4)\cos{kz}
\end{equation}
Thus $\omega t_2+\pi/4$ can be treated as a new phase factor $\phi$.
Also suggesting that the new $\hat{H}_{\pi/4}$ can be used in place of $\hat{H}_{\alpha}$. However there is another problem, we need to address, using $\omega t_{2}+\pi/4$ as $\phi$ or the new phase factor means $\phi$ is dependent upon $t_{2}$ which is very small, since we have assumed instantaneous transition, so that sudden approximation can be applied. However by small $t_{2}$, we mean $$t_{2}<\frac{\hbar \omega}{\Delta E}$$[8](assuming $t_{1}=0$), in our case, we performed Unitary transformations on the Hamiltonian $\hat{H}$, thus we had no change in the energy eigenvalues which means $\Delta E=0$. Thus $t_{2}$ can attain any real value of time, and we can still successfully apply the sudden approximation to our perturbative change. Thus $\phi$ can attain all the real values. Also the least value of phase factor $\phi$ is not just limited to $\pi/4$, since $-\pi/4$ is also a viable option for representing a Hamiltonian with pure uncertainties.Thus the phase factor$\phi$ can take any real value and is not constrained by any factor.

This paper also shows how fundamental is the Sudden approximation. We essentially replace one of the postulates of Copenhagen interpretation with Sudden Approximation. We also must focus on the fact that the sudden "collapse of states" was previously considered to be pure Quantum Mechanical Assumption. In this paper we had shown how it is nothing but a consequence of the Uncertainty principle which holds true for any canonical $\hat{p}$ and $\hat{q}$, for $\comm{\hat{p}}{\hat{q}}=i\hbar$
\section*{References}
[1] M.Planck, \textit{Verh.Dtsch.Phys.Ges.Berlin}\textbf{2},202(1990);\textit{ibid}. \textbf{2},237(1990) {\\}
[2]  C.L. Mehta and E.C.G.,\textit{Phys. Rev}. \textbf{138},B274(1965)
[3] E.C.G. Sudarshan, \textit{Phys. Rev. Lett.} \textbf{10},277(1963) {\\}
[4] C.L. Mehta, \textit{Phys. Rev. Lett} \textbf{18},752(1967) {\\}
[5]McCulloch, M.E. Gravity from the uncertainty principle. Astrophys Space Sci 349, 957–959 (2014).{\\}
[6]  W.P. Schleich \textit{Quantum optics in phase space}(Berlin: Wiley-VCH,2001) {\\}
[7]R.J. Glauber, \textit{Phys. Rev.} \textbf{131}, 2766(1963) {\\}
[8] Hitoshi Murayama, 221 A Lecture notes\\
http://hitoshi.berkeley.edu/221A/timedependent.pdf

\end{document}